# Plasmonics as a fabrication tool

Hamid T. Chorsi

Department of Electrical and Computer Engineering, University of California, Santa Barbara, CA 93106, USA

hamidchorsi@ucsb.edu

**Abstract**

Plasmonic nanolithography is a nanofabrication technique which exploits surface plasmons to generate subdiffraction patterns. In this review, we highlight the key concepts in plasmonic nanolithography, review the pioneering papers and provide a classification that can be useful in determining the proper method and tool for subdiffraction patterning. Considering the ongoing developments in the field of plasmonics, lithography using surface plasmons is an extremely promising approach for development of high-throughput nanofabrication techniques.

**Introduction**

Nanophotonics, the study of light at the nanoscale, has become a vibrant field of research, as the flow of light can be manipulated at length scales far below the optical wavelength, largely surpassing the classical limits imposed by diffraction. In plasmonic devices, free electrons in the metal can be excited by the electric component of light to have collective oscillations. Using metallic and dielectric nanostructures precisely sculpted into two-dimensional (2D) and 3D nanoarchitectures, light can be scattered, refracted, confined, filtered, and processed in fascinating new ways that are impossible to achieve with natural materials and in conventional geometries. Strictly speaking, plasmonics; investigates plasmons; quanta of collective oscillations of the conduction electrons in a plasma. Plasmonics can squeeze light into dimensions far beyond the diffraction limit by coupling light with the surface collective oscillation of free electrons at the interface of a metal and a dielectric. The physics of surface plasmons, is extremely interesting and heralds enchanting applications for nanofabrication. Recently, surface plasmon resonances (SPPs) have been used to fabricate nanostructures, especially for patterning nanoscale structures. The main purpose in this review is to highlight recent advances in surface plasmon-based nanofabrication, a great phenomenon that happens at the interface of plasmonic and micro/nano structures. We classify "Plasmonic nanolithography" into three categories: prism-coupled plasmonic nanolithography, grating-coupled plasmonic nanolithography, and direct writing plasmonic nanolithography.

**1. Prism-coupled Plasmonic nanolithography**

The idea of prism-coupled plasmonic nanolithography originated from the excitation of surface plasmons at the interface of a metal and dielectric via a prism. The main idea here is the use of evanescent waves (from the interaction of light with metal mask) to pattern photoresist. In 2006, Guo et al. proposed a large-area surface plasmon polariton interference lithography (LSPPIL) [1]. Figure 1(a), which is characterized by the Kretschmann configuration, shows the physical arrangement.

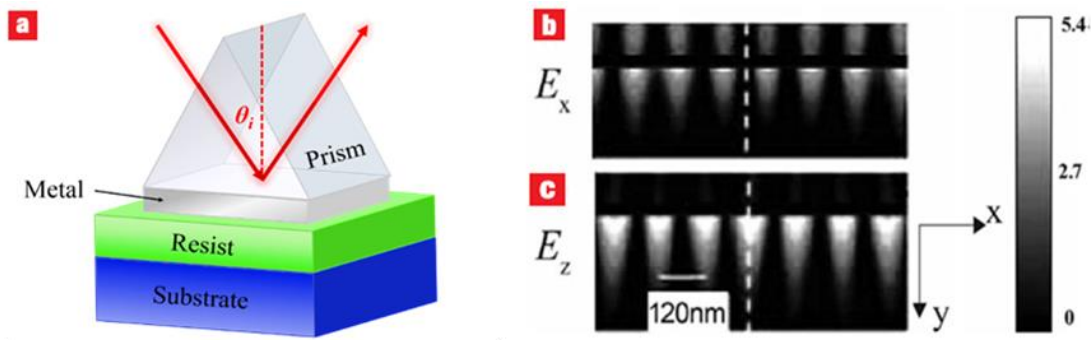

Figure 1. (a) Schematic of the LSPPIL process, (b-c) Electric field distribution of the interference patterns with TM incident wave.

An isosceles triangle is placed at the uppermost layer in order to excite the SPPs. The bottom surface of the prism is coated with a thin metal (silver) film and then brought into intimate contact with a thin photoresist coated on a substrate. When two mutually coherent TM (p-polarized) plane waves are incident on the base of the prism in the vicinity of the resonance angle, multiple counterparts of the SPPs arise everywhere on the interface. As a result, surface plasmon interference fringes are formed in the photoresist. Figure 1(b) shows the electric field distributions of the interference pattern with an incidence angle of 59.9 degrees at a wavelength of 441nm. Line-widths less than 65 nm has been achieved using this technique [2].

To extend and improve the performance and increase the flexibility of the prism-based SPPs interference, several implementation approaches have been demonstrated. Lim et al. proposed a thin-film patterning method based on the coupling between the surface plasmon mode and plasmon waveguide modes [3]. The crux of this method is resonance at three angles, 33.5, 56.3, and 90 degrees, respectively. Different SPPs interference patterns have been achieved using this method at different resonant angles.

Further improvement may be achieved by reducing the illumination wavelength. For reducing the feature size of a lithography pattern, authors in [4] have used 193 nm illumination wavelength in LSPPIL and replaced the silver coating by tungsten. Future sizes less than 32 nm will likely be attained using the proposed method.

In classical prism-coupled plasmonic nanolithography (contact based), the photoresist layer and the metallic thin film are placed in intimate contact; this may result in damage or pollution of surfaces. It is worth mentioning that He *et al*. in 2010 [5] proposed a backside exposure method to fabricate nanostructures that can prevent the drawbacks of the contact exposure scheme. Schematic of the backside-exposure SPPIL structure is shown in Figure 2(a). The structure includes a prism, matching fluid layer and glass substrate, silver film, a resist layer, and an air layer.

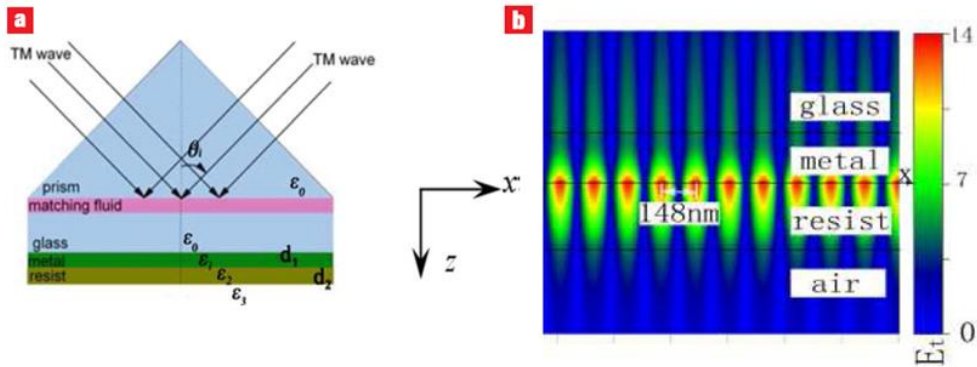

Figure 2. (a) Backside-exposure SPPIL structure, (b) Simulation result by FDTD, with incident wavelength 441.6nm, thickness of silver film 40nm, refractive index of resist 1.53 and thickness 50nm. Electric field distribution when refractive index of the prism is 1.89.

In summary, prism-coupled plasmonic nanolithography has several advantages, including masklessness, high transmission, soft contact, and large-area fabrication capability.

## 2. Grating-coupled Plasmonic nanolithography

Grating-coupled plasmonic nanolithography uses metallic grating masks along with appropriate structures to excite SPPs and pattern nanoscale features. As distinct from a Kretschmann scheme, the mask grating based scheme is much more compact. The schematic of plasmonic lithography configuration using metal mask is shown in Figure 3(a). It consists of a metal mask, which can be fabricated on a thin quartz glass by electron-beam lithography and lift off process. The mask is brought into intimate contact with a photoresist coated on a silica substrate. Normally incident light tunnels through the mask via SPPs and reradiates in to the photoresist.

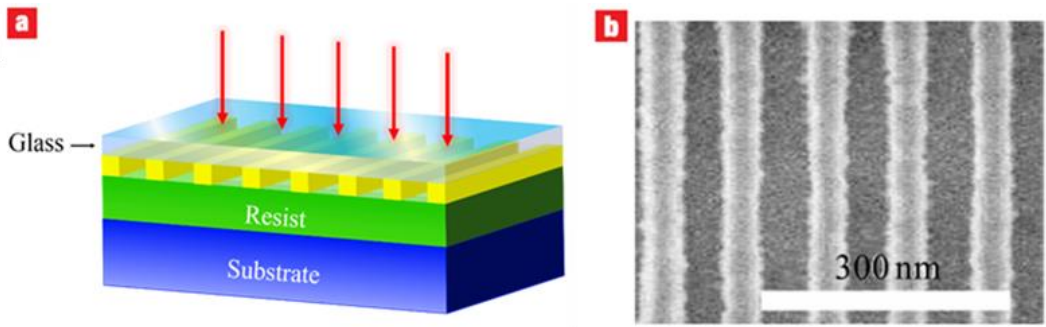

Figure 3. (a) Schematic of a single metallic grating lithography. (b) SEM picture of the pattern using a single metallic grating lithography.

In 2004, Luo *et al*. [6] performed analytical and numerical calculations to demonstrate this method. Figure 4 shows the considered nano-pattern. The proposed structure can be easily experimentally realized on a 2-mm thick quartz by electron beam lithography. To generate a high-contrast interference pattern, Doskolovich *et al*. optimized the diffraction grating for the substrate with a metal film at the interface of dielectric material. They have used a 3D metal–dielectric diffraction structure to generate the 2D surface plasmon interference patterns seen in Figure 4 [7].

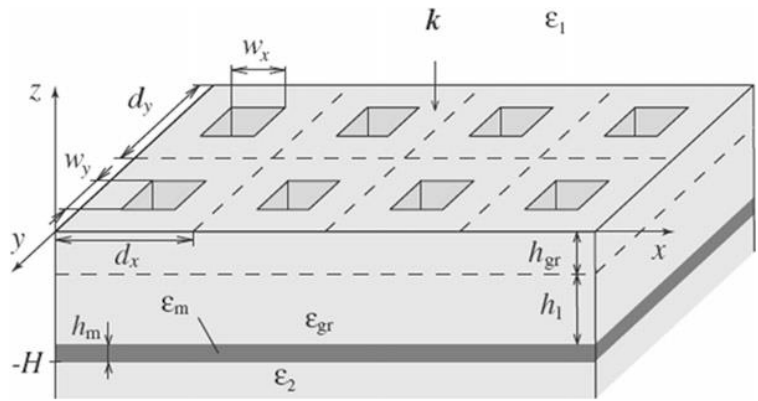

Figure 4. Geometry of a 2D surface plasmon interference structure [7].

Figure 5 shows an interference pattern generated underneath the metal film in the case of TM wave incident with a wavelength of 550 nm. ψ is the polarization angle between the direction of the electric field vector $E$ and the x-axis.

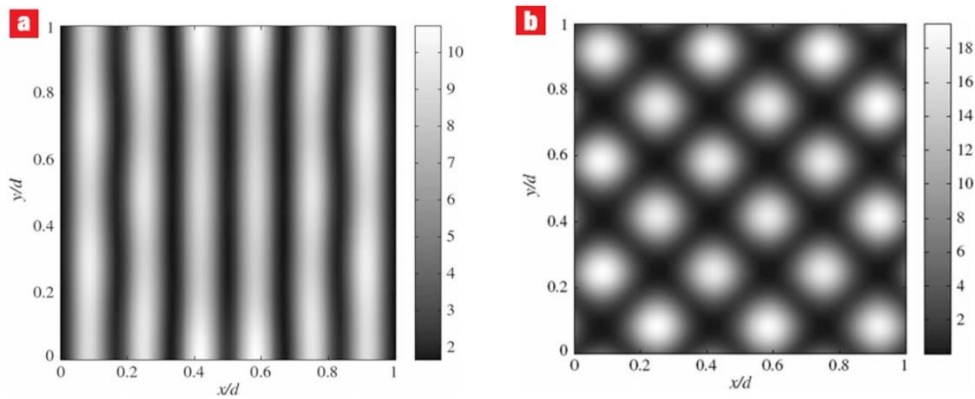

Figure 5. (a) Electric field distribution underneath the metal film with TM wave incidence (ψ=0 degrees). (b) Electric field distribution in the case of linear polarization wave (ψ=45 degrees).

Patterns with even smaller sizes have been achieved using surface relief metal grating [8]. To further improve the intensity and quality of the interference fringes, surface plasmon interference lithography (SPIL) assisted by a Fabry–Perot (F-P) cavity was proposed by Liang et al. in 2015 [9]. The SPIL process is presented in Figure 6(a).

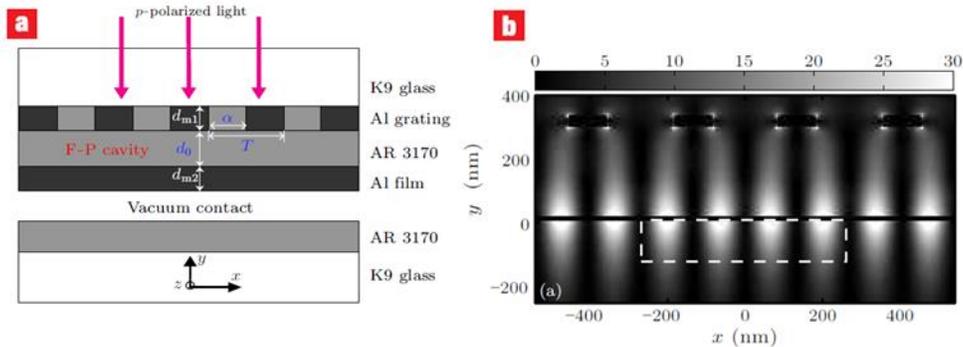

Figure 6. (a) Fabry–Pérot cavity with the same photoresist material, this is useful to improve the intensity of interference fringes in the resist. Vacuum contact between two parts can be achieved with the assistance of a contact mask aligner.

One key challenge in using plasmonic nanolithography is that the evanescent plasmons can only generate patterns at the surface of photoresist (usually a few nanometers). For patterning thicker photoresist, plasmons need to penetrate deeper. Utilizing grating-coupled approach, recent work by Jin et al. [10] exploits concentric annular grating to realize deep subwavelength lithography.

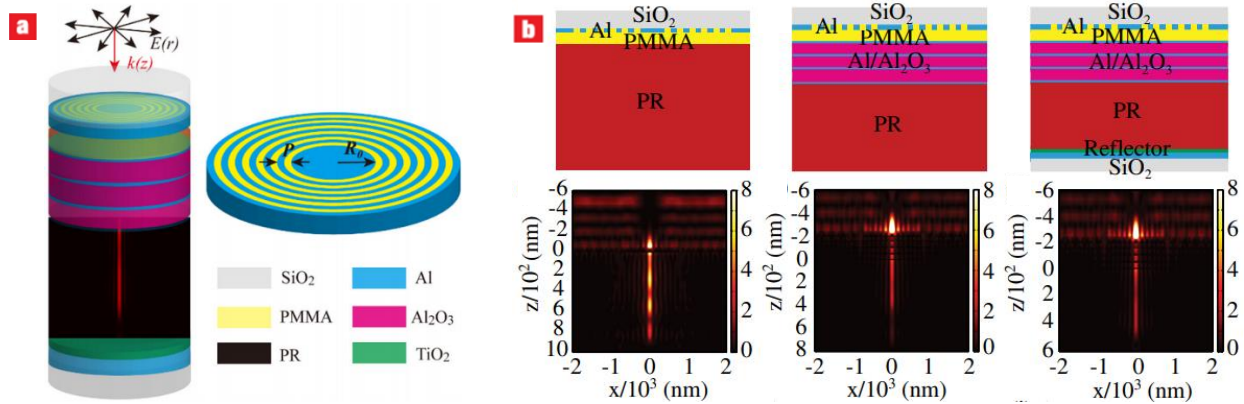

Figure 7. Deep subwavelength lithography. (a) Schematic of concentric grating on top of hyperbolic metamaterial. (b) Intensity distributions inside photoresist for three different photolithography systems [10].

Obtained results show that a focal depth of over 500 nm can be realized using this technique. Hyperbolic metamaterial made of $Al/Al_2O_3$ stacks is placed beneath the grating to focus high-k diffracted beams into the substrate.

## 3. Focused plasmonic nanolithography

Focused plasmonic nanolithography, which is a promising technique due to its sub-diffraction resolution limit, can be described as an extension of scanning probe lithography (SPL). In this technique, resist is scanned and illuminated by a plasmonic lens-induced super-focused light spot.

In 2008, Wang et al. demonstrated a novel, practical plasmonic near-field scanning optical microscopy system (NSOM) experimentally for near-field lithography [11]. A subwavelength aperture is coated with thin Al and placed at the apex of the cone for the formation of a conic plasmonic lens. Subsequently, a high-speed maskless nanolithographic approach with a plasmonic lens was proposed by Srituravanich et al. [12] to improve efficiency. This idea combines the plasmonic lens with a flying head to focus the surface plasmon wave on a substrate rotating with a high speed. Its working principle is schematically shown in Figure 8.

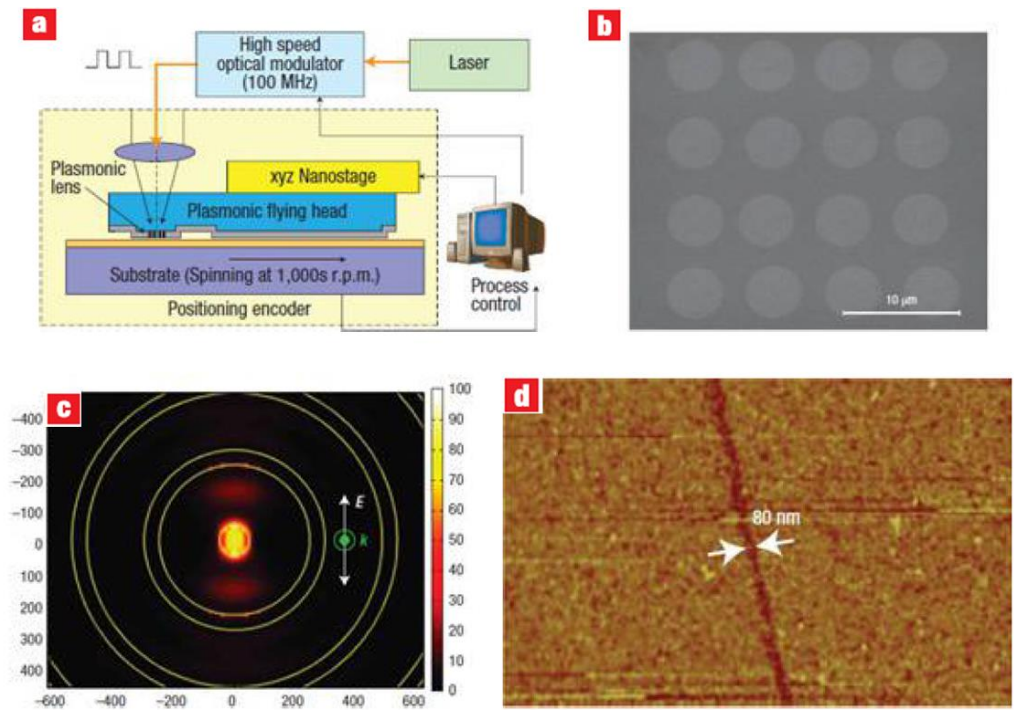

Figure 8. Maskless focused lens nanolithography with plasmonic lens array. (a) Schematic of the main process and the experimental setup. (b) SEM image of an array of plasmonic lenses fabricated on an air bearing surface. (c) FEM simulation of the focused spot using plasmonic lens. (d) AFM image of a resist pattern with 80 nm line width on a photoresist.

In this lithography experiment, a UV continuous-wave laser with 365 nm was focused down to a spot of several micrometers onto a plasmonic lens. It is further focused by the plasmonic lens to a spot of sub-100 nm to expose the spinning disk for writing arbitrary patterns on a thermal photoresist. The laser pulses were controlled by an electro-optic modulator according to the signals from a pattern generator. A "plasmonic flying head" with arrays of plasmonic lenses fabricated on its bottom surface was employed as the direct writing probe. Figure 8(b) shows the SEM image of the fabricated plasmonic lens, which has a 4 by 4 array of concentric ring gratings with a through hole perforated at the center.

Among all nanoparticles, the bowtie structure is attractive because of its triangular geometry, which leads to the "lightning-rod" effect at the gap apexes [13-15]. In 2009, Kim et al. reported a plasmonic lithography technique with bowtie-shaped contact probes [16].

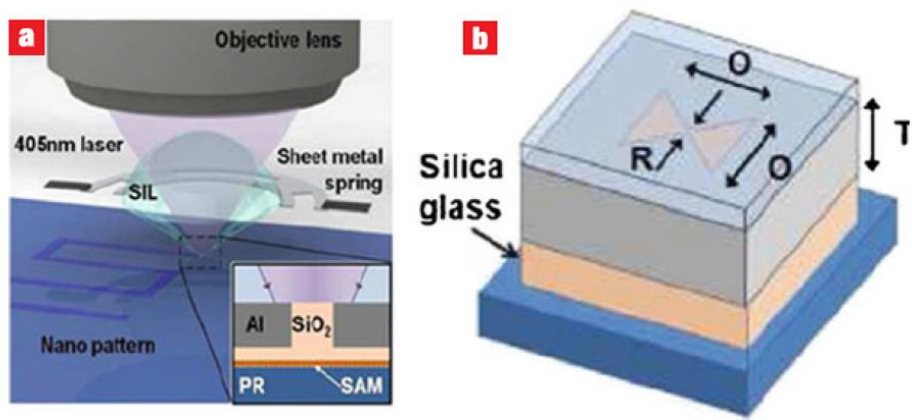

Figure 9. (a) The structure of the bowtie-shaped probe. (b) Schematic of a bowtie aperture.

As shown in Figure 9, first an Al film of 120 nm thickness is coated on the flat surface of a conically shaped lens with a flat surface in 30 μm. In order to ensure the adhesion of the Al film and the glass substrate, a Cr layer of 2 nm thickness is added. Then, FIB is employed to fabricate bowtie apertures with about 140 nm dimension in order to achieve the highest transmission of the aperture. Finally, a single-atom self-assembled film was coated on the surface of silica film to reduce the friction between the probe and the photoresist during scanning. The experiment results are shown in Figure 10; the smallest line width of 50 nm was realized when the polarized laser light had parameters of 405 nm wavelength and 0.5 mW power.

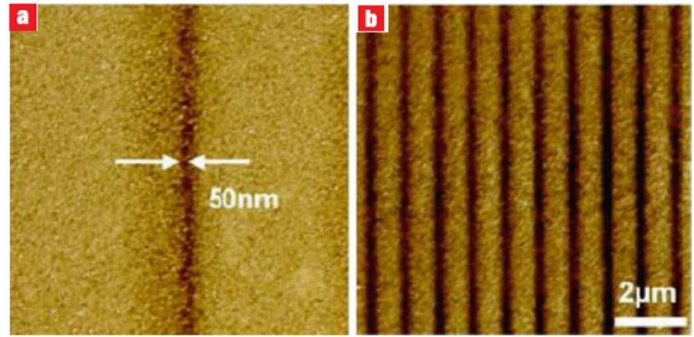

Figure 10. (a) A single line of 50 nm width. (b) multiple line pattern of 150 nm width with 1 μm pitch.

Recently, efforts have been redoubled to decrease the dimensions and increase the throughput of focused plasmonic nanolithography. Authors in [17] have utilized a bowtie aperture combined with a metal-insulator-metal (MIM) scheme to obtain sub-32 nm (λ/12) high-aspect plasmonic spots. In order to improve the throughput, a record number of near-field optical elements, an array of 1,024 bowtie antenna apertures, were simultaneously employed to generate a large number of patterns by carefully controlling their working distances over the entire array with an optical gap metrology system [18].

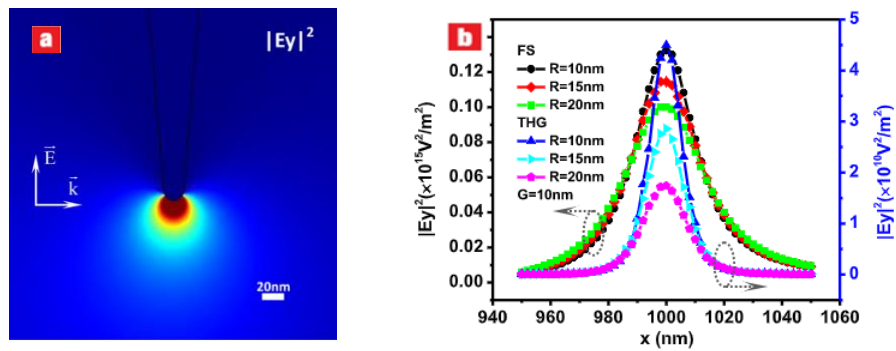

Figure 11. NSOM lithography (a) localized surface plasmons at the apex of a gold tip. (b) Full width at half maximum (FWHM) of the first and third harmonic signal intensity [19].

Over 100-time increase in electric field intensity can be achieved via localizing surface plasmons at the apex of an NSOM. This approach has been theoretically proposed to produce features less than 10 nm [19]. From Figure 11, it can be seen that the obtained third-harmonic can be focused at the tip with extremely small full width at half maximum.

In conclusion, plasmonic nanolithography has great potential to be integrated with current nano/micro fabrication techniques to realize sub-diffraction patterns, structures and eventually optoelectronic devices. The main focus of this review was to classify and highlight some of the main advances in the area of plasmonic nanolithography and encourage future advances in this field.